\documentclass[conference]{IEEEtran}
\ifCLASSINFOpdf
\else
\fi
\usepackage{amsmath}
\usepackage{amssymb}
\usepackage{amsfonts}
\usepackage{graphicx}
\usepackage{subfig}
\usepackage{algorithm}
\usepackage{algorithmic}
\usepackage{multirow}
\usepackage{cite}

\newtheorem{proposition}{Proposition}
\newtheorem{theorem}{Theorem}
\newtheorem{lemma}{Lemma}

\newtheorem{remark}{Remark}

\begin{document}
%
\title{Traffic-Aware Transmission Mode Selection in D2D-enabled Cellular Networks with Token System}

\author{
\IEEEauthorblockN{Yiling Yuan\IEEEauthorrefmark{1}, Tao Yang\IEEEauthorrefmark{1}, Hui Feng\IEEEauthorrefmark{1}, Bo Hu\IEEEauthorrefmark{1}\IEEEauthorrefmark{2}, Jianqiu Zhang\IEEEauthorrefmark{1}, Bin Wang\IEEEauthorrefmark{1} and Qiyong Lu\IEEEauthorrefmark{1}}
\IEEEauthorblockA{
\IEEEauthorrefmark{1} Research Center of Smart Networks and Systems, School of Information Science and Engineering\\
\IEEEauthorrefmark{2}Key Laboratory of EMW Information (MoE)\\
Fudan University, Shanghai, China, 200433\\
Emails: \{yilingyuan13, taoyang, hfeng, bohu, jqzhang01, wangbin, lqyong\}@fudan.edu.cn}}
\maketitle

\begin{abstract}
We consider a D2D-enabled cellular network where user equipments (UEs) owned by rational users are incentivized to form D2D pairs using tokens. They exchange tokens electronically to ``buy'' and ``sell'' D2D services. Meanwhile the devices have the ability to choose the transmission mode, i.e. receiving data via cellular links or D2D links. Thus taking the different benefits brought by diverse traffic types as a prior, the UEs can utilize their tokens more efficiently via transmission mode selection. In this paper, the optimal transmission mode selection strategy as well as token collection policy are investigated to maximize the long-term utility in the dynamic network environment. The optimal policy is proved to be a threshold strategy, and the thresholds have a monotonicity  property. Numerical simulations verify our observations and the gain from transmission mode selection is observed.
\end{abstract}


%
\IEEEpeerreviewmaketitle

\section{Introduction}
To meet the dramatically increasing traffic demand and provide better user experience, the device-to-device (D2D) communication has been proposed recently. This technology, which enables direct communication between two mobile users in proximity, has attracted attention in both industry and academic \cite{22803, 5350367, 6805125}. The adoption of D2D communication allows high-rate, low-delay and low-power transmissions\cite{6163598}.
\par
Recent researches on D2D communication mainly focus on developing various optimization and game theory frameworks for mode selection, resource allocation or interference management in order to maximize throughput or to improve energy efficiency \cite{5910123, 7174559, 6766256}. These studies are based on the assumption that there are many devices already in D2D communication mode. However, this assumption needs to be re-examined in realistic scenarios. The UEs are possessed by self-interested users who aim to maximize their individual utilities. In practice, they would have no incentive to provide D2D service unless receiving satisfactory rewards. Therefore, it is crucial to design a proper incentive mechanism to encourage UEs to form D2D pairs\cite{7166194}.
\par
In this paper, we design a token-based incentive system. In such system, UEs pay tokens to or gain tokens from other UEs in exchange for D2D service. Some previous works have investigated the token system on cooperative relaying in cellular networks \cite{6544193,7181721}. In \cite{6544193}, authors designed a token incentive scheme from the perspective of a system designer. In \cite{7181721}, authors investigated how UEs can learn the token gathering strategies online. However, neither of them takes into account how UEs make decisions when facing two alternatives, i.e. D2D link versus cellular link. The former one has to consume tokens while the latter one does not. In practice, there are various types of traffic which will result in different benefits in D2D communication. If the decision on transmission mode selection is considered, tokens can be utilized more efficiently. Intuitively, the UE could spend more tokens on more beneficial traffic types to improve his utility. Therefore, it is crucial to answer the question ``when to use tokens'' or ``which transmission mode to choose'' equivalently. To the best of our knowledge, our work is the first attempt in literature  to investigate token consuming policy in the token system designed for D2D-enabled cellular networks.
\par
 In this paper, we consider a D2D-enabled cellular network where UEs are incentivized to form D2D pair using tokens. We formulate a Markov decision process (MDP) model to characterize the interaction between each UE and environment. i.e. transmission mode selection policy and token collection strategy. When traffic arrives, a UE needs to first choose the transmission mode, and then determines whether to accept D2D request if idle. The objective of a UE is to maximize his long-term utility, which is defined as the difference between the benefit he obtains when receiving data through D2D link and the cost he pays when providing D2D service. Furthermore, the structure of the optimal policy is investigated. Unlike\cite{7181721,7070657}, the optimal policy is analytically proved to be threshold in the number of the tokes instead of just taking this property as an assumption. Moreover, it turns out that the threshold increases as a function of the benefits of the traffic types. The numerical simulations verify our observations and the gain from transmission mode selection is observed.
\par
The rest of this paper is organized as follows. In Section II, the system model is discussed. In Section III, the MDP model for individual UE's decision problem is developed. In Section IV, we investigate the structure of the optimal policy. Section V gives some numerical simulation results, and finally section VI concludes this paper.
\section{System Model}
\subsection{Network Model}
In this paper, we consider a D2D-enabled wireless cellular network and the slot based system is adopted. At each slot, when traffic arrives, such as transferring a file, the UE will choose the transmission mode and start a transmission procedure. The transmission modes include cellular mode and D2D mode. The former mode corresponds to  the conventional cellular communication and the latter mode represents D2D communication. According to the given policy as well as the available information, the decision is made at the beginning of each slot. Note that each slot may last several seconds and each type of traffic may last multiple slots.
\par
Without loss of generality, we assume that for any type, D2D mode can always obtain higher benefit than cellular mode. It is reasonable due to lower power consumption and higher throughput of D2D link. Suppose the utility for cellular mode is $0$ for convenience. Considering different requirements for different traffic types, we define the specific utility for each type of traffic according to its characteristics. There are some widely used classification in literature under various practical consideration. Basically, in terms of throughput requirements, traffic can be divided into two types, stream traffic and elastic traffic \cite{5738225}. Besides, traffic can also be classified as video traffic, audio traffic and file transfer according to the types of applications\cite{5466675}.
\par
Similar to \cite{7018008}, we do not specify a concrete traffic classification. Instead, we assume there are $N$ types of traffic and the traffic type set is denoted as $\mathcal{S}^o=\{s_1,s_2,\cdots,s_N \}$. Especially, we regard $s_0$ as a special type of traffic, namely, the idle state. Hence we can define the extended traffic type set as $\mathcal{S} = \mathcal{S}^o\cup \{s_0\}$. The stationary probability of each type $s\in\mathcal{S}$ is $p(s)$ with $0<p(s)<1$ and $\sum_{s\in\mathcal{S}}p(s)=1$. We denote $b_s$ as the benefit of D2D mode for traffic type $s\in\mathcal{S}^o$. Moreover, we assume that $0<b_{s_1}<b_{s_2}<\cdots<b_{s_N}$.
\subsection{Token System}
Although D2D communication has multiple advantages, the UEs are generally reluctant to provide D2D service since this incurs cost and provides them with no reward. To overcome this difficulty, we use token system to incentive UEs to accept D2D requests. Specifically, a UE must spend tokens in exchange for receiving data through D2D link, and can only earn tokens by providing D2D service for other UEs. Because the device works in half-duplex mode and the traffic demand must be met, so it is reasonable to assume that only in idle state, a UE can provide D2D transmission service.
\par
The token system has many advantages\cite{6544193}. First, there is no extra payment exchange involved, which avoids many financial problems associated with other monetary incentive schemes. Second, no personal information exchange is required, which allows secure implementation. Recently, several techniques which could enable electronic token transaction, have been proposed\cite{buttyan2001nuglets}. We assume that our token system is implemented using such technologies.
\par
The entire system is described as follows. Consider UE $j$ decides to start a D2D transmission. At this point, UE $j$ sends a D2D request to the selected UE $j^*$ under a predefined criterion. If the request is accepted, UE $j$ will pay one token to UE $j^*$ through token exchange system. Otherwise, the UE $j$ will seek another UE to forward his traffic. If there is no UE accepting the request, UE $j$ has to deliver the data through BS. Moreover, we assume that the UE can serve only one UE simultaneously. Therefore, an idle UE will choose only one or reject all when he receives multiple D2D requests.

\section{Problem Formulation}
In this section, we formulate the optimal policy for a UE based on MDP model. When a UE has no token, he has no choice but to choose cellular mode. In addition, a UE would spend as many tokens as possible on the traffic types with high utility in order to maximize his utility. Therefore, it is needed to investigate the optimal strategy, which includes transmission mode selection policy and token collection strategy.
\subsection{State and Action Spaces}
\textbf{Token holding state:} At any given slot $t$, the UE holds $k_t\in\mathcal{K}=\{0,1,\cdots,K\}$ tokens, where $K$ is the maximal number of tokens allowed in the system.
\par
\textbf{Traffic type state:} At different slots, the UE may have different type of traffic. Denote the type of traffic in slot $t$ as $s_t\in\mathcal{S}$. Assume that the traffic types of different slots are independent mutually.
\par
The state parameters defined above can be used to describe the UE's private information at slot $t$. Hence, let $\Omega_{t}=(s_t,k_t)$ denote the state of the UE at slot $t$.
\par
When $s\neq s_0$, which means a specified traffic arrives, the UE can take an action to choose D2D mode or cellular mode. We denote the action taken when $s\neq s_0$ as $a_M\in A_M=\{0,1\}$. $a_M=0$ and $a_M=1$ represent the cellular mode and D2D mode, respectively.
\par
When $s=s_0$, the UE can decide whether to accept D2D requests from other UEs. In this situation, we denote the action taken as $a_R\in A_R=\{0,1\}$. $a_R=0$ is the action that the UE chooses to accept the D2D request to earn one token, and $a_R=1$ represents the action that the UE refuses to provide D2D service for other UEs. Putting all these together, the action space $A(s,k)$ is shown in Table.\ref{tab1}.
\begin{table}
\centering
\caption{Action spaces}
\label{tab1}
  \begin{tabular}{|c|c|c|c|}
    \hline
    State $(s,k)$   & Action space & Action & Physical meanings \\
    \hline
    \multirow{2}*{$s\neq s_0$}  & \multirow{2}*{$A_M$}  & $a_M=0$ & choose cellular mode \\
    \cline{3-4}
                            &                       & $a_M=1$ & choose D2D mode \\
        \hline
    \multirow{2}*{$s=s_0$}  & \multirow{2}*{$A_R$}  & $a_R=0$ & accept any D2D request \\
    \cline{3-4}
                            &                       & $a_R=1$ & refuse any D2D request \\
    \hline
  \end{tabular}
\end{table}

\subsection{Transition Probability}
Now we discuss the state transition probability. Let $P\{(s',k')|(s,k),a\}$ denote the state transition probability function, which represents the probability that the UE transfers from state $\Omega=(s,k)$ to state $\Omega'=(s',k')$ depending on the action $a$.
\par
Because the D2D request may not be accepted and a UE may not receive any D2D requests even if he takes the action $a_R=0$, the state transition is influenced by the complicated varying environment. We use a stochastic model to describe the environmental dynamics. The associated environmental factors are shown in Table.\ref{tab2}. Specifically, we use $0<p<1$ to denote the probability of receiving D2D requests when the UE takes the action $a_R=0$, and use $0<q<1$ to denote the probability of the D2D request being accepted when the UE takes the action $a_M=1$. These parameters are unknown a priori, but can be learned from history or other reinforcement learning methods\cite{kaelbling1996reinforcement}, such as Q-learning.
\begin{table}[t]
\centering
\caption{Environmental factors}
\label{tab2}
\begin{tabular}{|c|c|}
\hline
Parameters & Physical meanings \\
\hline
$p$ & probability of receiving D2D requests \\
\hline
$q$ & probability of its D2D request being accepted \\
\hline
\end{tabular}
\end{table}
\par
Consequently, the state transition probability is presented in (\ref{equ1}). We will explain it in detail later.

\begin{equation}
\label{equ1}
\begin{split}
  &P\{(s',k')|(s,k),a\}=\\
  &
  \begin{cases}
  p(s')\{(1-a_M)+a_M(1-q)\} &s \ne {s_0},k>0,k' = k \\
  p(s')q{a_M}               &s\ne {s_0},k>0,k' = k-1\\
  p(s')                     &s \ne {s_0},k = 0,k' = k\\
  p(s')\{a_R+(1-a_R)(1-p)\} &s = {s_0},k < K,k' = k\\
  p(s')p{(1-a_R)}           &s = {s_0},k < K,k' = k + 1\\
  p(s')                     &s = {s_0},k = K,k' = k\\
  0                         &\text{otherwise}
  \end{cases}
\end{split}.
\end{equation}
\par
At first, we consider the case in which $s\neq s_0$ thus $a=a_M\in A_M$. If $k>0$, the number of tokens can decrease by one or stay unchanged depending on the selected action. There are two possibilities for the transition from $(s,k)$ to $(s',k)$: the first situation is that the UE takes the action $a_M =0$, indicating $P\{(s',k)|(s,k),a_M=0\}=p(s')$; the other one suggests that the D2D request is rejected by all potential UEs while the UE takes the action $a_M=1$, and it corresponds to $P\{(s',k)|(s,k),a_M=1\}=p(s')(1-q)$. Therefore, we obtain $P\{(s',k)|(s,k),a_M\}=p(s')\{(1-a_M)+a_M(1-q)\}$ when putting them together. Transition from $(s,k)$ to $(s',k-1)$ will happen only when the UE takes the action $a_M=1$ and the D2D request is accepted. Thus, the transition probability is $p(s')qa_M$, which means that the transition is possible only when $a_M=1$. Besides, the probability of transition from $(s,0)$ to $(s',0)$ is $p(s')$ no matter which action is taken since the action $a_M=1$ is meaningless here. Otherwise the transition probability is zero. Following the similar argument, we can get the transition probability when the UE is idle.

\subsection{Reward}
When the UE provides D2D service for another UE, the cost incurred is defined as $c$. The cost can be thought as the average cost of all possible D2D transmissions because we only care about the average utility in our model. Thus, we can get the expected reward $\mu(s,k,a)$ depending on state $(s,k)$ and action $a$ as follows.
\begin{equation}
\label{equ2}
\mathbb{E}\{\mu(s,k,a)\}=
\begin{cases}
  -cp(1-a_R)        &s=s_0 \\
  qa_Mb_sI(k>0) &s\neq s_0
\end{cases}.
\end{equation}
where $\mathbb{E}\{\cdot\}$ is the expectation and $I(\cdot)$ is the indicator function.

\subsection{Optimization Problem Formulation}
A policy $\pi$ is defined as a function to specify the action $\pi(s,k)$ to be taken for the state $(s,k)$. When $s=s_0$, $\pi(s,k)$ represents the transmission mode selection policy and it corresponds to token collection policy when $s \neq s_0$. The expected utility obtained by executing policy $\pi$ starting at state $(s_0,k_0)$ is given by
\begin{equation}
V^\pi(s_0,k_0) = \mathbb{E}\{\sum_{t=0}^\infty \beta^t\mu (s_t,k_t,\pi(s_t,k_t))\},
\end{equation}
where $\beta\in(0,1)$ is the discounted factor.
\par
Our goal is to find the optimal policy $\pi^*$ to maximize the expected utility, which can be expressed as the optimization problem shown in (\ref{equ4}).
\begin{equation}
\label{equ4}
\pi^*=arg\max_\pi V^\pi(s_0,k_0).
\end{equation}
Value iteration or policy iteration\cite{bertsekas2007dynamic} can be used to obtain the optimal policy when $p$ and $q$ are known. When these environmental parameters are unknown, Q-learning\cite{kaelbling1996reinforcement} can be adopted.
\section{Optimal Policy for a Single UE}
In this section, we investigate the structure of optimal policy. We will prove that the optimal policy is threshold. In \cite{6544193}, this property is proved only for one-dimensional state case, but a two-dimensional state case is analyzed here.
\par
Let $V^*(s,k)=V^{\pi^*}(s,k)$ for brevity. It is given by the solution of Bellman equation shown in (\ref{equ5})\cite{bertsekas2007dynamic}.
\begin{equation}
\label{equ5}
\begin{split}
  &V^*(s,k)=\\
  &\max_{a\in A(s,k)} \left\{ \mathbb{E}\{\mu(s,k,a)\} +\beta \sum_{s'\in\mathcal{S}}p(s',k'|s,k,a){V^*}(s,k)  \right\}.
\end{split}
\end{equation}
The optimal policy $\pi^*(s,k)$ is the action $a\in A(s,k)$ to maximize the right hand side of Bellman equation. It is easy to find out that $\pi^*(s,0)=0(s\neq s_0)$ and $\pi^*(s_0,K)=1$. From the Bellman equation, it turns out that the optimal strategy has the following one-shot deviation property\cite{6544193}.
\begin{lemma}
The optimal strategy $\pi^*$ has following property:
\begin{itemize}
\item[(1)] For $s\neq s_0,k>0$, $\pi^*(s,k)=0$ if and only if
\begin{equation}
\label{equ6}
\beta \sum_{s'\in\mathcal{S}} p(s')\left\{ {{V^*}(s',k) - {V^*}(s',k - 1)} \right\}  \geq {b_s}.
\end{equation}
\item[(2)] For $s=s_0,k<K$, $\pi^*(s,k)=0$ if and only if
\begin{equation}
\label{equ7}
  \beta \sum_{s'\in\mathcal{S}} p(s')\left\{ {{V^*}(s',k+1) - {V^*}(s',k)} \right\}  \geq c.
\end{equation}
\end{itemize}
\end{lemma}
\begin{IEEEproof}
For $s\neq s_0,k>0$, based on Bellman equation (\ref{equ5}), $\pi^*(s,k)=0$ if and only if
\begin{equation*}
  \begin{split}
  & qa_Mb_s + \beta \sum_{s'\in\mathcal{S}}p(s',k'|s,k,a_M){V^*}(s,k) \Big|_{a_M = 0} \geq  qa_Mb_s +\\
  & \beta \sum_{s'\in\mathcal{S}}p(s',k'|s,k,a_M){V^*}(s,k) \Big|_{a_M = 1}.
  \end{split}
\end{equation*}
Using the transition probability in (\ref{equ1}), we can obtain the following inequality.
\begin{equation*}
  \begin{split}
  \beta \sum_{s'\in\mathcal{S}} p(s')V^*(s',k)\geq qb_s +  \beta \sum_{s'\in\mathcal{S}} p(s') \{ & (1-q)V^*(s',k)  + \\
                                                                                                       & qV^*(s',k-1) \}.
  \end{split}
\end{equation*}
After some simple algebraic operations, inequality (\ref{equ6}) can be verified.
\par
Following the similar argument, we can prove the second part of Lemma 1.
\end{IEEEproof}
\par
The LHS of (\ref{equ6}) is the opportunity cost for using one token at this point and the RHS of (\ref{equ6}) is the immediate utility brought by this action. Since the opportunity cost is higher than the immediate utility, the UE will choose $a_M=0$, namely cellular mode. We can interpret (\ref{equ7}) in a similar way.
\par
Here we assume that the the environmental factors $p$ and $q$ are known as a prior. Thus the value iteration algorithm can be used to obtain the optimal policy, which is depicted in Algorithm 1.

\begin{algorithm}[t]
\caption{\textbf{Value Iteration Algorithm}}
\label{alg1}
\textbf{Initialize:} $V^0(s,k)=0,\forall s\in \mathcal{S},\forall k \in \mathcal{K}$\\
\textbf{Loop:}
\begin{itemize}
\item[1]\textbf{Update the policy} $\{\pi^{n+1}(s,k)\}$:\\
Set $\pi^{n+1}(s,0)=0(s\neq s_0)$ and $\pi^{n+1}(s_0,K)=1$\\
(1) For $s\neq s_0$, if
\begin{small}
\begin{equation*}
\beta \textstyle{\sum_{s'\in\mathcal{S}}} p(s')\left\{ {{V^n}(s',k) - {V^n}(s',k - 1)} \right\}  \geq {b_s}.
\end{equation*}
\end{small}
then $\pi^{n+1}(s,k)=0$ and $\pi^{n+1}(s,k)=1$ otherwise.\\
(2) For $s = s_0$, if
\begin{small}
\begin{equation*}
\beta \textstyle{\sum_{s'\in\mathcal{S}}} p(s')\left\{ {{V^n}(s',k+1) - {V^n}(s',k)} \right\}  \geq c.
\end{equation*}
\end{small}
then $\pi^{n+1}(s,k)=0$ and $\pi^{n+1}(s,k)=1$ otherwise.
\item[2]\textbf{Update the utility function} $\{V^{n+1}(s,k)\}$:
\begin{equation*}
\begin{split}
V^{n+1}(s,k)=&\mathbb{E}\{\mu(s,k,\pi^{n+1}(s,k))\} +\\
  &\beta \sum_{s'\in\mathcal{S}}p(s',k'|s,k,\pi^{n+1}(s,k)){V^n}(s,k)
\end{split}
\end{equation*}
\end{itemize}
\textbf{Until:} $\max_{s,k}|V^{n+1}(s,k)-V^{n}(s,k)|<\epsilon$
\end{algorithm}
\par
Now we show the marginal decrease of the utility function $V^{n}(s,k)$ at each iteration of Algorithm 1. This property is depicted in Theorem 1 in detail.
\begin{theorem}[The marginal diminishing utility]
At each iteration of Algorithm 1, the following inequality holds:
\begin{equation}
\label{equ8}
  V^{n}(s,k+1)-V^{n}(s,k)\leq V^{n}(s,k)-V^{n}(s,k-1), n\geq 0.
\end{equation}
\end{theorem}
\begin{IEEEproof}
We will use induction to show that (\ref{equ8}) holds for $n\geq 0$.
  \par
  1) Due to the initiation step of Algorithm 1, (\ref{equ8}) holds for all $n=0$.
  \par
  2) Suppose the induction hypothesis holds for some $n\geq 0$. In order to prove (\ref{equ8}) holds for $n+1$, the proof includes two parts. At first we will show that $\pi^{n+1}(s,k)$ has threshold structure, which will be used to verify (\ref{equ8}) for $n+1$ in the second part. And for the sake of notational conciseness, we define $\Delta^n(k)\triangleq \sum_{s'\in\mathcal{S}} p(s')V^n(s',k+1)$, and then the following inequality holds by using the induction hypothesis:
  \begin{equation}
  \label{equ9}
  \Delta^{n}(s,k+1)-\Delta^{n}(s,k)\leq \Delta^{n}(s,k)-\Delta^{n}(s,k-1).
  \end{equation}
  \par
  We first show the threshold structure of $\pi^{n+1}(s,k)$. It suffices to prove that if $\pi^{n+1}(s,k+1)=0$, then $\pi^{n+1}(s,k)=0$. When $s\neq s_0$, given the step 2 of the algorithm and using (\ref{equ9}), we get the inequality $\Delta(k)-\Delta(k-1)\geq \Delta(k+1)-\Delta(k)\geq b_s$, so $\pi^{n+1}(s,k)=0$. Similarly, we can prove it when $s=s_0$.
  \par
  Next we will prove that given the utility function obtained in step 2 of the algorithm, (\ref{equ8}) holds for $n+1$.
  \par
  When $s\neq s_0$, we only need to consider four cases due to the threshold structure of the policy.
  \par
   Case 1: $\pi^{n+1}(s,k-1)=\pi^{n+1}(s,k)=0$ and $\pi^{n+1}(s,k+1)=1$. Thus
   \begin{align*}
     &V^{n+1}(s,k-1)=\Delta^n(k-1),\\
     &V^{n+1}(s,k)=\Delta^n(k),\\
     &V^{n+1}(s,k+1)=qb_s+q\Delta^n(k)+(1-q)\Delta^n(k+1).
   \end{align*}
   Then, we can get:
   \begin{align*}
        &V^{n+1}(s,k+1)-V^{n+1}(s,k)\\
   =    &b_sq + (1-q)\{ \Delta(k+1)-\Delta(k)\}\\
   \overset{(a)}{\leq} &q\{ \Delta(k)-\Delta(k-1)\}+(1-q)\{ \Delta(k+1)-\Delta(k)\}\\
   \leq &\Delta(k)-\Delta(k-1)\\
   =    &V^{n+1}(s,k)-V^{n+1}(s,k-1).
   \end{align*}
  Using the fact that $\pi^{n+1}(s,k)=0$ amounts to $\Delta^n(k)-\Delta^n(k-1)\leq b_s$, we can obtain inequality (a).
  \par
   Case 2: $\pi^{n+1}(s,k-1)=0$ and $\pi^{n+1}(s,k)=\pi^{n+1}(s,k+1)=1$. Thus
   \begin{align*}
     &V^{n+1}(s,k-1)=\Delta^n(k-1),\\
     &V^{n+1}(s,k)=b_s+q\Delta^n(k-1)+(1-q)\Delta^n(k),\\
     &V^{n+1}(s,k+1)=qb_s+q\Delta^n(k)+(1-q)\Delta^n(k+1).
   \end{align*}
   Then, the following inequality can be obtained:
   \begin{align*}
        &V^{n+1}(s,k)-V^{n}(s,k-1)\\
   =    &qb_s-q\{\Delta(k)-\Delta(k-1)\}+\{\Delta(k)-\Delta(k-1)\}\\
   \overset{(a)}{\geq} &\Delta(k)-\Delta(k-1).
   \end{align*}
   Inequality (a) holds because when $\pi^{n+1}(s,k)=1$, then $\Delta^n(k)-\Delta^n(k-1)\geq b_s$. Moveover, we can find out that:
   \begin{align*}
        &V^{n+1}(s,k+1)-V^{n}(s,k)\\
   =    &q{\Delta(k)-\Delta(k-1)}+(1-q){\Delta(k+1)-\Delta(k)}\\
   \leq &\Delta(k)-\Delta(k-1).
   \end{align*}
Therefore, it is obvious that (\ref{equ8}) holds for $n+1$ in this situation.
  \par
  For the case where $\pi^{n+1}(s,k-1)=\pi^{n+1}(s,k)=\pi^{n+1}(s,k+1)=0$ or $\pi^{n+1}(s,k-1)=\pi^{n+1}(s,k)=\pi^{n+1}(s,k+1)=1$, it is easy to verify the inequality.
  \par
  Similarly, we can verify the inequality $V^{n+1}(s,k+1)-V^{n+1}(s,k)\leq V^{n+1}(s,k)-V^{n+1}(s,k-1)$ when $s=s_0$.
\end{IEEEproof}
\par
\begin{remark}
Theorem 1 indicates that the marginal reward of owning an additional token decreases. The incentive of holding a token is that the UE can use the token to request D2D service to improve his utility. However, keeping tokens has inherent risk modeled by $\beta$, which exponentially ``discounts'' future rewards.
\end{remark}
\par
Furthermore, from the proof of Theorem 1, we can find an important fact that the optimal policy is a threshold strategy in $k$ for a given traffic type.
\begin{proposition}[Threshold structure]
The optimal policy is a threshold strategy when the traffic type is given. Specifically, there exits a constant $K_{th}(s)$ depending on the type of traffic $s\in \mathcal{S}$, such that:
  \begin{equation}
    \pi^*(s,k) =
    \begin{cases}
      0 & k < K_{th}(s) \\
      1 & k \geq K_{th}(s)
    \end{cases}.
  \end{equation}
\end{proposition}
\begin{IEEEproof}
  We only consider the case when $s\neq s_0$ here and the proof is similar when $s=s_0$. Recall that $\pi^*(s,0) = 0(s\neq s_0)$, therefore it is sufficient to show that if $\pi^*(s,k) = 1(k\geq 1,s\neq s_0)$, then $\pi^*(s,k+1) = 1$.
  \par
  Suppose $\pi^*(s,k) = 0(s\neq s_0)$. According to Lemma 1, we find out that
  \begin{equation*}
  \beta \sum_{s'\in\mathcal{S}} p(s')\left\{ {{V^*}(s',k) - {V^*}(s',k - 1)} \right\}  \leq {b_s}.
  \end{equation*}
  Additionally, Theorem 1 implies that ${V^*}(s',k+1 ) - {V^*}(s',k) \leq {V^*}(s',k ) - {V^*}(s',k - 1)$. Therefore, we have
  \begin{equation*}
  \beta \sum_{s'\in\mathcal{S}} p(s')\left\{ {{V^*}(s',k+1) - {V^*}(s',k)} \right\}  \leq {b_s},
  \end{equation*}
  which implies that $\pi^*(s,k+1) = 1$.
\end{IEEEproof}
\par
Intuitively, for traffic type $s\neq s_0$, if the UE chooses D2D mode when owning $k$ tokens, he is more likely to still choose D2D mode when more tokens is available. In fact, many research works make this assumption due to its simplicity when they build their models. Unlike these works, we analytically prove that optimal policy has a threshold structure instead of just assuming this property without rigorously proving its optimality.
\begin{remark}
According to Proposition 1, only $|\mathcal{S}|$ thresholds is needed to define the optimal policy. Therefore, the size of search space would be significantly reduced due to the small amount of traffic types. Note that this property still holds when the traffic types of adjacent slots are dependent.
\end{remark}
\par
Moreover, it turns out that the thresholds have a monotonicity property.
\begin{proposition}[Monotonicity]
If $b_{i}<b_{j}(i,j\neq s_0)$, then $K_{th}(j)\leq K_{th}(i)$ where $K_{th}(s)$ is the threshold defined in Proposition 1.
\end{proposition}
\begin{IEEEproof}
It is sufficient to verify that if $b_i<b_j(i,j\neq s_0)$ and $\pi^*(j,k)=0$, then $\pi^*(i,k)=0$. According to Lemma 1, we can find out that $\sum_{s'\in\mathcal{S}} p(s'){V^n(s',k)-V^n(s',k-1)}\geq b_j \geq b_i$, and thus we can get $\pi^*(i,k)=0$ using the sufficient condition for the optimal policy.
\end{IEEEproof}
\par
Proposition 2 implies that the more beneficial traffic types have higher probability to be served in D2D mode due to the lower threshold. It means that the UE will spend more tokens on those traffic types. Consequently, the UE's long-term utility is improved.
\section{Numerical Simulations}
In this section, we give simulations to verify the analyzed results. At first, we present several numerical results to show the structure of the optimal policy and illustrate the behavior of the optimal threshold $K_{th}(s)(s\neq s_0)$ with respect to other parameters. We assume that $s_1,s_2,s_3,s_4$ belongs to $\mathcal{S}^o$ and $p_{s_0}=p_{s_1}=p_{s_2}=p_{s_3}=p_{s_4}=0.2$. The benefits of these traffic types are $3,4,5,6$, respectively, the cost $c=1$ and $K=20$. These parameters are set for illustration purpose, and a more realistic scenario will be considered later.
\begin{figure}
  \centering
  \includegraphics[width=2.1 in]{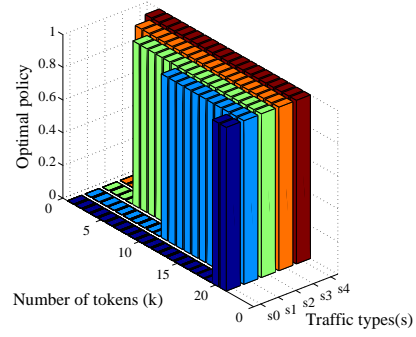}\\
  \caption{Structure of the optimal policy with $\beta=0.99, p=0.5, q=0.5$.}\label{OptimalPolicy}
\end{figure}

\begin{figure}[!t]
\centering
\subfloat[\scriptsize{Optimal thresholds with different $\beta$}]{\includegraphics[width=1.7in]{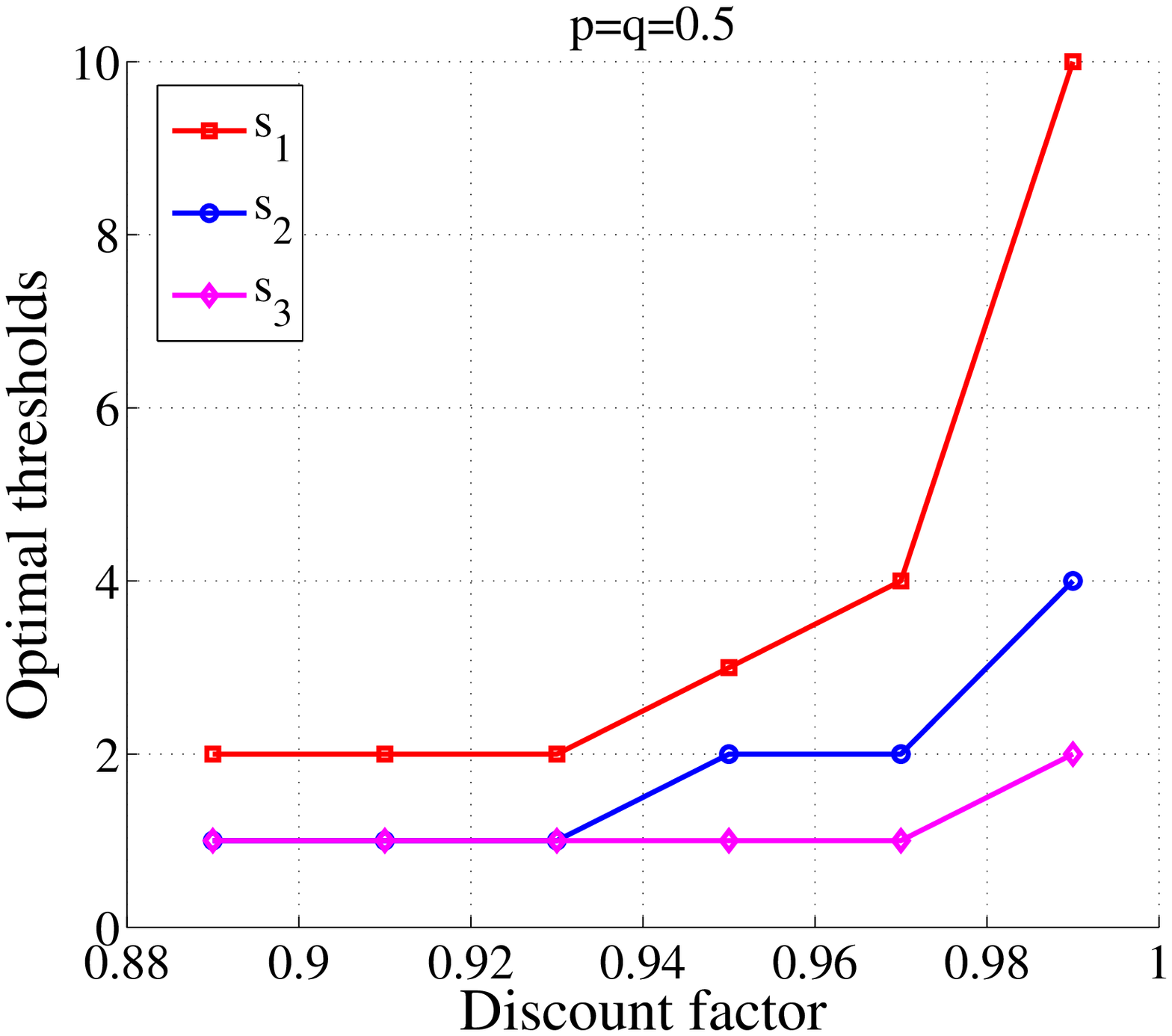}%
\label{Thresholds_beta}}
\hfill
\subfloat[\scriptsize{Optimal thresholds with different $p$}]{\includegraphics[width=1.7in]{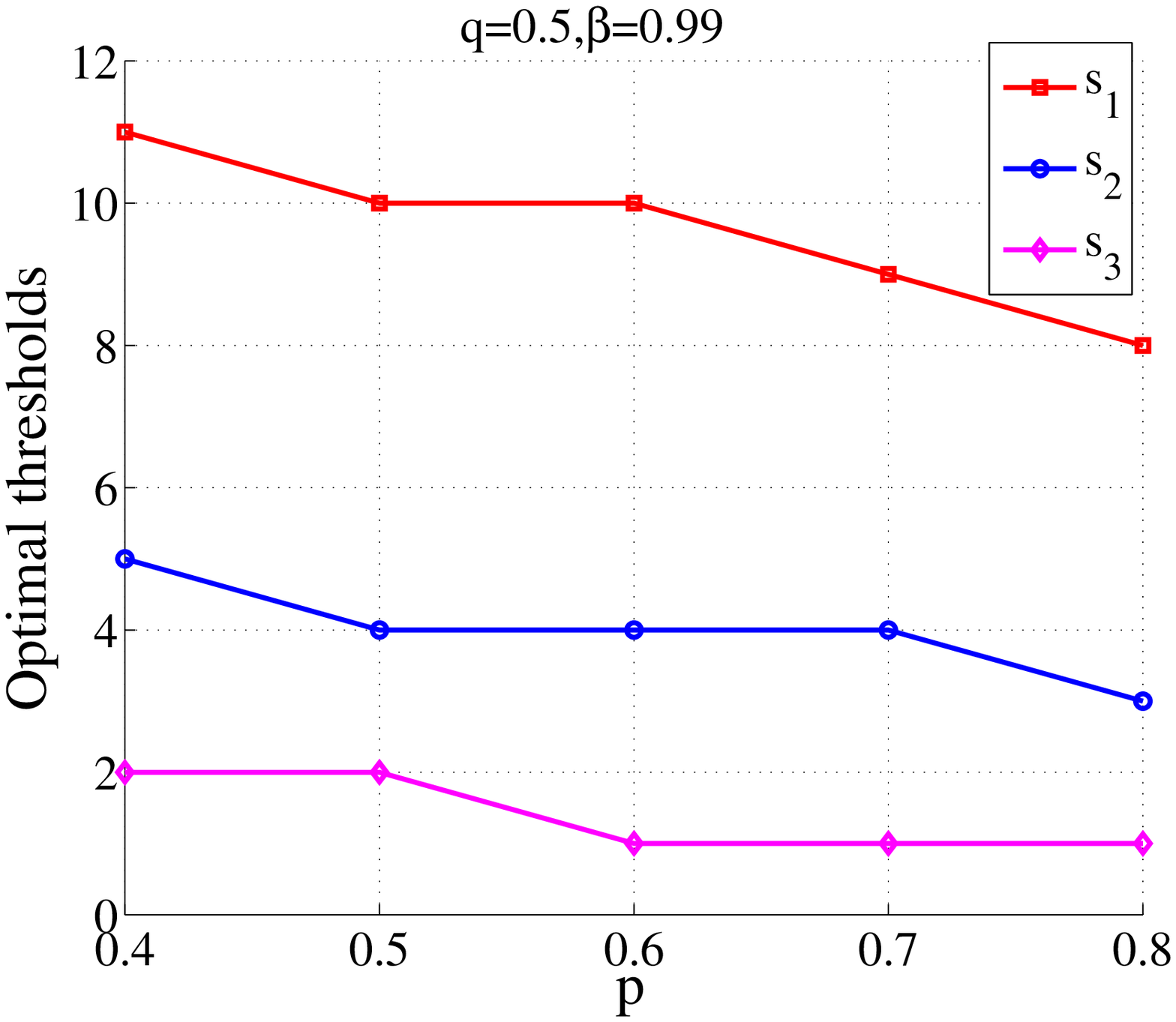}%
\label{Thresholds_p}}
\hfill
\subfloat[\scriptsize{Optimal thresholds with different $q$}]{\includegraphics[width=1.7in]{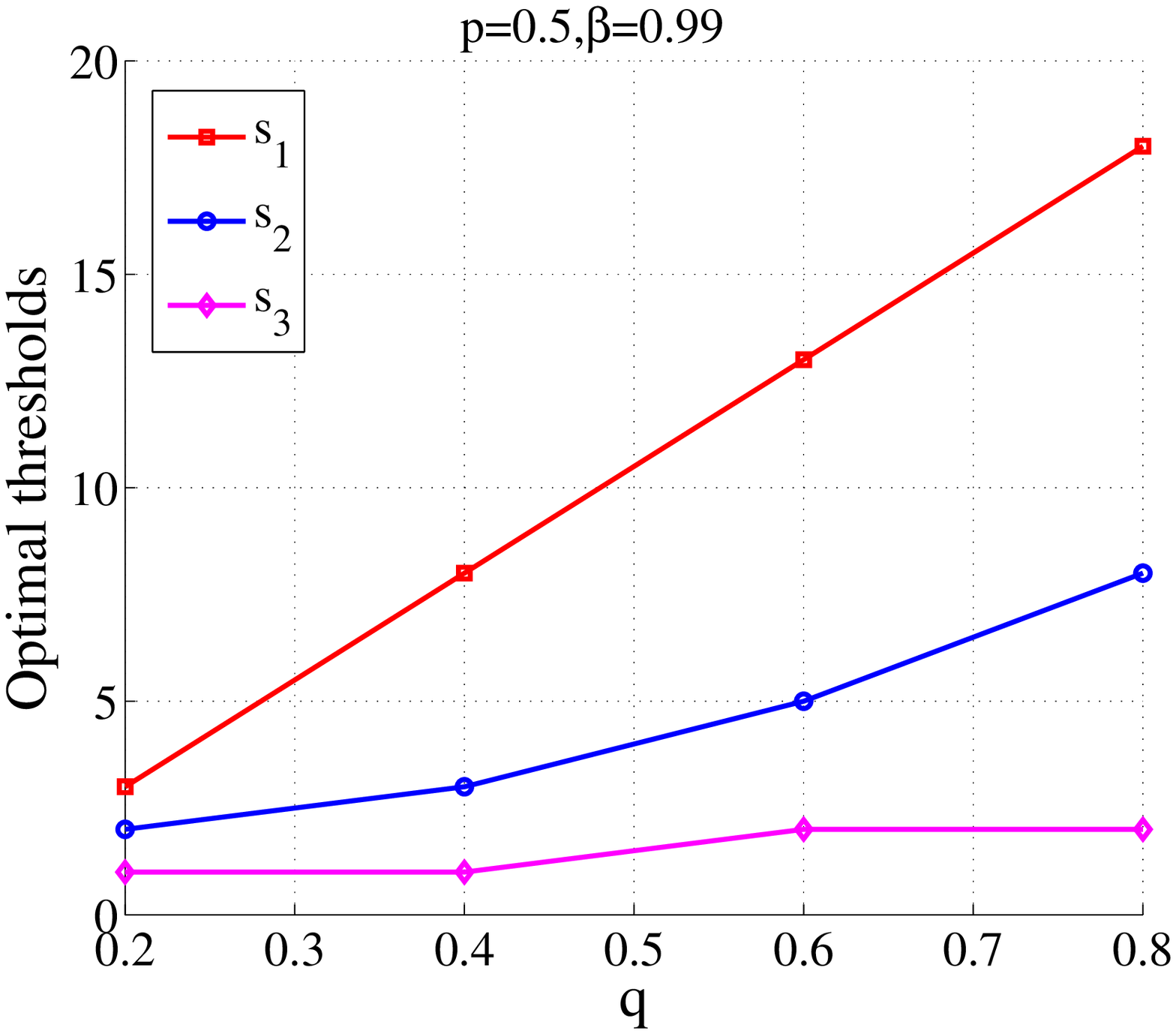}%
\label{Thresholds_q}}
\caption{Thresholds with different parameters}
\label{Thresholds}
\end{figure}
\par
 The optimal policy is given Fig.\ref{OptimalPolicy}. As indicated in Proposition 1, the optimal policy is threshold in the token state $k$. Fig.\ref{Thresholds_beta} shows the thresholds vary with respect to the discount factor $\beta$. Note that the threshold for the most beneficial traffic type is always one, which is omitted here. The optimal threshold is non-decreasing in $\beta$. It is because, if a UE is far-sighted, he intends to wait for a better chance to consume tokens, i.e. more beneficial traffic types, and thus has less incentive to use tokens if the benefit of the current traffic is low. Fig.\ref{Thresholds_p} illustrates the thresholds vary with respect to the environmental factor $p$. The optimal threshold decreases as $p$ increases. This happens allowing for being easier to collect tokens as $p$ increases, which leads to more incentive for the UE to use tokens albeit $b_s$ is low. Fig.\ref{Thresholds_q} shows the variation of the thresholds with respect to the environmental factor $q$. The optimal threshold decreases as $q$ decreases, since that the D2D request is seldom accepted when $q$ is low, and thus a UE has more incentive to take every opportunity to seek D2D service. Additionally, as proved in Proposition 2, the threshold decreases with the increase in benefit $b_s$.
\par
Furthermore, we give simulations to show the gain obtained from transmission mode selection. A more realistic scenario is considered, where traffic is divided into two types: $s_v$-video traffic and $s_e$-elastic traffic. Thus the extended traffic type set is denoted as $\mathcal{S}=\{s_0,s_v,s_e\}$. The mean opinion score
(MOS) is often used as a subjective measure of the network quality in literature. The benefit of each traffic type in our simulation is defined as the difference in the MOS obtained by two transmission modes. The MOS estimations of two traffic types depend on experienced Peak Signal-to-Noise-Ratio (PSNR) $P_{snr}$ and throughput $\theta$, respectively. They are expressed as follows\cite{7018008}:
\begin{align}
&Q_{s_v}(P_{snr})=4.5-\frac{3.5}{1+\exp(b_1(P_{snr}-b_2))},\\
&Q_{s_e}(\theta)=b_3\log(b_4\theta),
\end{align}
where $b_1=1$, $b_2=5$, $b_3=2.6949$ and $b_4=0.0235$. In our simulations, $P_{snr}=10\text{db}$ and $\theta=1500\text{kbps}$ for D2D mode. Meanwhile, $P_{snr}=5\text{db}$ and $\theta=1000\text{kbps}$ for cellular mode. Moreover, the stationary probability is set as $p_{s_0}=0.3$, $p_{s_v}=0.2$ and $p_{s_e}=0.5$. Let the environmental factors $p=q=0.8$ and they are known as a prior. The discount factor $\beta$ is set to be $0.99$ and the cost $c$ is set to be 0.4. The simulation runs $10^6$ slots.
\par
\begin{figure}[!t]
\centering
\subfloat[\scriptsize{Token usage distribution}]{\includegraphics[width=1.7in]{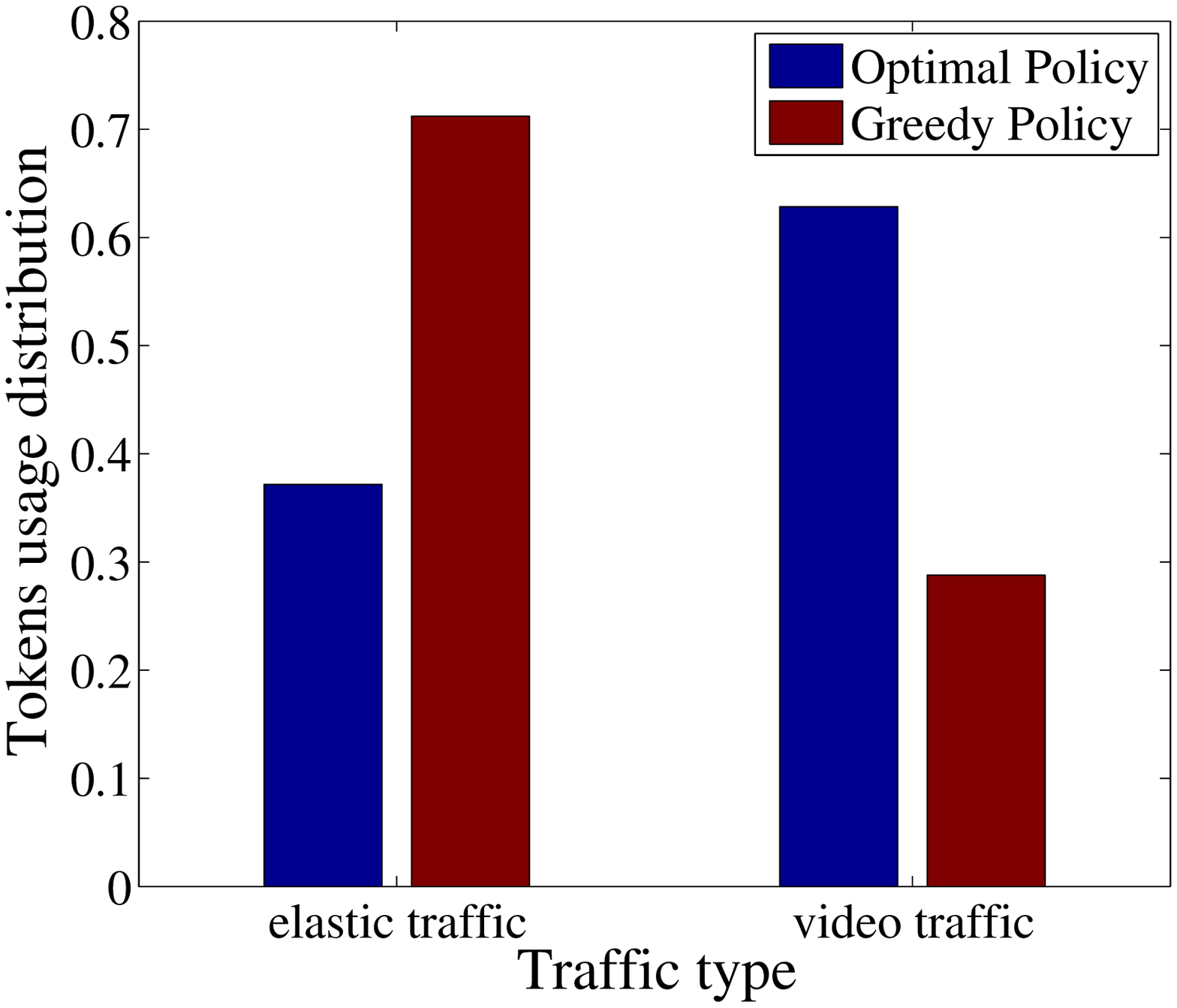}%
\label{CompareTokenUsage}}
\hfill
\subfloat[\scriptsize{Average utility comparison}]{\includegraphics[width=1.7in]{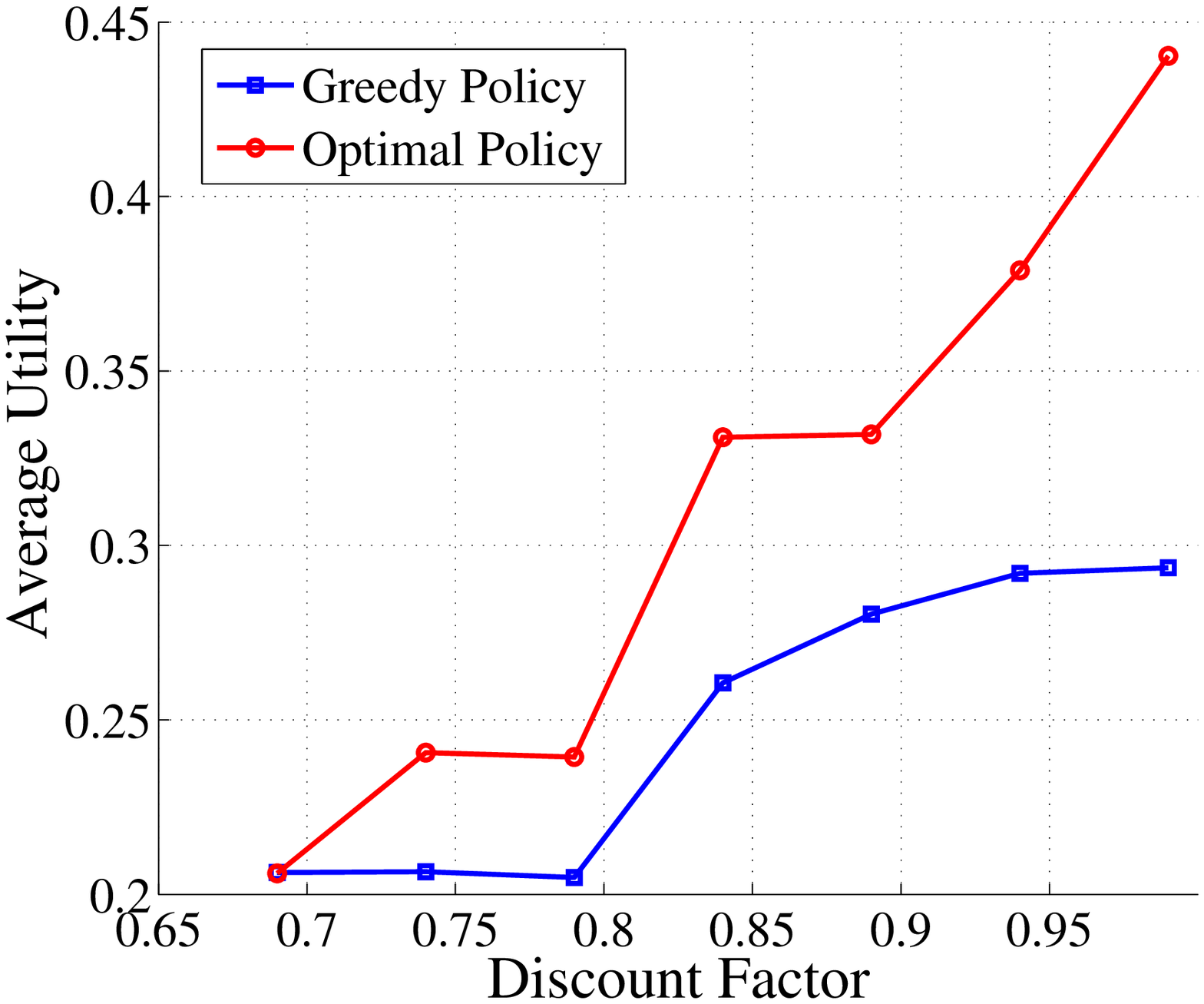}%
\label{CompareBeta}}
\caption{Performance Comparison}
\label{PerformanceCompare}
\end{figure}

A greedy policy is considered for comparison. We assume that the UE will choose D2D mode when having any tokens, and the goal of this policy is to optimize the token collection strategy only. Fig.\ref{CompareTokenUsage} shows the distribution of token usage over different traffic types. We can find out that the distribution is proportional to $p_s$ when the greedy policy in executed. In contrast, since the different benefits of different traffic types are distinguished, more tokens are spent on the more beneficial traffic types and the number of tokens spent on the least beneficial traffic type $s_e$ dramatically decreases. Fig.\ref{CompareBeta} presents the average utilities of two policy with different discount factor $\beta$. The utilities of both policy increase with increasing $\beta$ due to the fact users with higher $\beta$ are more far-sighted. Moreover, the gain obtained by considering transmission mode selection can be observed. However, the gap tends towards zero when $\beta$ is small. That's because the UE with low $\beta$ is myopia so that he inclines to spend token no matter the traffic type is, which is similar to the greedy policy. Besides, the emergence of plateau of the curves is because the variation of $\beta$ is not large enough to change the policy.

\section{Conclusion}
In this paper, we consider a D2D-enabled cellular network where selfish UEs are incentivized to form D2D pairs using tokens. We formulate a MDP model to characterize UE's behavior including transmission mode selection strategy as well as token collection policy. Moreover, we prove that the optimal strategy is threshold in the token state and show that the threshold increases as a function of the benefits related to the the traffic types. In our future work, we will explore the optimal selection of the maximum number of tokens so that the incentive mechanism can approach the altruism mechanism.

\bibliographystyle{IEEEtran}
\bibliography{IEEEabrv,reference}

\end{document}